\documentclass[prd,nofootinbib]{revtex4}
\usepackage{graphicx}
\usepackage{latexsym}

\begin{document}

\title{Constraining Slow-Roll Inflation in the Presence of Dynamical Dark Energy}

\author{Jun-Qing Xia}
\author{Xinmin Zhang}

\affiliation{Institute of High Energy Physics, Chinese Academy of
Science, P.O. Box 918-4, Beijing 100049, P. R. China}

\date{\today}

\begin{abstract}

In this paper we perform a global analysis of the constraints on
the inflationary parameters in the presence of dynamical dark
energy models from the current observations, including the
three-year Wilkinson Microwave Anisotropy Probe (WMAP3) data,
Boomerang-2K2, CBI, VSA, ACBAR, SDSS LRG, 2dFGRS and ESSENCE (192
sample). We use the analytic description of the inflationary power
spectra in terms of the Horizon-flow parameters $\{\epsilon_i\}$.
With the first order approximation in the slow-roll expansion, we
find that the constraints on the Horizon-flow parameters are
$\epsilon_1<0.014~(95\%~C.L.)$ and
$\epsilon_2=0.034\pm0.024~(1\sigma)$ in the $\Lambda$CDM model. In
the framework of dynamical dark energy models, the constraints
become obviously weak, $\epsilon_1<0.022~(95\%~C.L.)$ and
$\epsilon_2=-0.006\pm0.039~(1\sigma)$, and the inflation models
with a ``blue" tilt, which are excluded about $2\sigma$ in the
$\Lambda$CDM model, are allowed now. With the second order
approximation, the constraints on the Horizon-flow parameters are
significantly relaxed further. If considering the non-zero
$\epsilon_3$, the large running of the scalar spectral index is
found for the $\Lambda$CDM model, as well as the dynamical dark
energy models.

\end{abstract}



\maketitle

\section{Introduction}

Inflation in the very early universe is the most attractive
paradigm, which is driven by a potential energy of a scalar field
called inflaton and its quantum fluctuations turn out to be the
primordial density fluctuations which seed the observed large
scale structures (LSS) and the anisotropies of cosmic microwave
background radiation (CMB). Inflation theory has successfully
passed several non-trivial tests. The current cosmological
observations are in good agreement with an adiabatic and scale
invariant primordial spectrum, which is consistent with single
field slow-roll inflation predictions. And the large angle
anti-correlation is found in the temperature-polarization power
spectrum, which is the signature of adiabatic superhorizon
fluctuations at the time of decoupling \cite{WMAP1}.

In $1998$, the analysis of the redshift$-$distance relation of
type Ia supernova (SNIa) revealed the existence of the another
stage of accelerated expansion that started rather recently when a
mysterious new energy component dubbed dark energy (DE) dominated
the energy density of the Universe \cite{Riess}. The nature of
dark energy is among the biggest problems in modern physics and
has been studied widely. The simplest candidate of dark energy is
the cosmological constant (CC) however it suffers from the
fine-tuning and coincidence problems \cite{CCproblem}. To
ameliorate these dilemmas some dynamical dark energy models such
as Quintessence \cite{Quintessence}, Phantom \cite{Phantom} and
K-essence \cite{kessence}. Given our ignorance of the nature of
dark energy, constraining the evolution of DE the equation of
state (EoS) by cosmological observations is of great significance.
Interestingly, there exists some hints that the EoS of dark energy
has crossed over $-1$ at least once from current astronomical
observations \cite{Feng:2004ad,huterer}, namely Quintom dark
energy model, which greatly challenges the above mentioned dark
energy models.

In $2006$, the WMAP group \cite{wmap3:2006:1} obtained the
constraint on the scaler spectral index $n_s=0.958\pm0.016$, which
deviates from the simple scale-invariant primordial spectrum and
disfavors the inflationary models with a ``blue" tilt at more than
$2\sigma$. Alternatively, the scale-invariant
Harrison-Zel'dovich-Peebles (HZ) spectrum ($n_s=1,r=0$) is
disfavored about $3\sigma$ \cite{wmap3:2006:1}. And the large
running of the scalar spectral index is still allowed
\cite{wmap3:2006:1,Easther:2006tv}. It seems that the
scale-invariant spectrum is disfavored and the dynamics of
Inflation has been detected. Similar results have also been found
in the literature from the current observational data
\cite{Otherwork}. But it's noteworthy that these analysis are
based on the $\Lambda$CDM model. In the framework of dynamical
dark energy models, the constraints on the inflationary parameters
can be relaxed due to the degeneracy among the inflation and dark
energy parameters \cite{Xia:Inf1,Xia:Inf2}.

In this paper we use the current cosmological observations to
carry out a first detailed study on the inflationary parameters in
terms of the Horizon-flow parameters $\{\epsilon_i\}$ in the
presence of dynamical dark energy models. Our results show that
the dynamics of dark energy models weaken the constraints on the
Horizon-flow parameters significantly.


\section{Method and Data}

In order to compare the theoretical predictions of inflation
models with the cosmological observations, we often parameterize
the primordial power spectra of scalar and tensor perturbations
as:
\begin{eqnarray}
P_s(k)&=&A_s \exp\left[(n_s-1)\ln
\left(\frac{k}{k_{\ast}}\right)+\frac{\alpha_s}{2}\ln^2\left(\frac{k}{k_{\ast}}\right)\right]~,\nonumber\\
P_t(k)&=&A_t \exp\left[n_t\ln
\left(\frac{k}{k_{\ast}}\right)+\frac{\alpha_t}{2}\ln^2\left(\frac{k}{k_{\ast}}\right)\right]~,\label{paraPS}
\end{eqnarray}
where $n_i$ is the spectral index, $\alpha_i$ denotes the running
of the spectral index and $k_{\ast}$ is the pivot scale. In this
paper we use the analytic description of the inflationary power
spectra in terms of the Horizon-flow parameters
$\{\epsilon_i\}$\footnote{In the literature \cite{Peiris:2006sj}
the slow-roll parameters, $\epsilon$, $\eta$ and $\xi$, are also
used to constrain the inflation models.}, which are based on the
Hubble parameter during inflation and its derivatives, defined as
\cite{Schwarz:2001vv}:
\begin{equation}
\epsilon_1=-\frac{\dot{H}}{H^2}~,~\epsilon_{i+1}=\frac{d\ln{|\epsilon_i|}}{dN}=\frac{\dot{\epsilon_i}}{H\epsilon_i}~~(i\geq1)~,\label{HFF}
\end{equation}
where $N$ is the number of e-foldings. Following the
Eq.(\ref{paraPS}), the power spectrum can be obtained as an
expansion of the power spectrum in terms of the logarithmic
wavenumber \cite{SecOrder}:
\begin{equation}
\ln\frac{P(k)}{P_0(k)}=b_0+b_1\ln\left(\frac{k}{k_{\ast}}\right)+\frac{b_2}{2}\ln^2\left(\frac{k}{k_{\ast}}\right)+\cdots~,\label{expansion}
\end{equation}
where $P_{s0}=H^2G/\pi\epsilon_1$, $P_{t0}=16H^2G/\pi$ and the
coefficients $b_i$ given in Ref.\cite{SecOrder} are related to the
Horizon-flow parameters
$\{\epsilon_i\}$\footnote{Ref.\cite{Casadio:2006wb} used the
method of comparison equations in the study of the cosmological
perturbations and obtained the similar coefficients $b_i$.}. The
coefficients for the scalar spectrum are:
\begin{eqnarray}
b_{s0}&=&-2(C+1)\epsilon_1-C\epsilon_2+\left(\frac{\pi^2}{2}-2C-7\right)\epsilon^2_1
      +\left(\frac{7\pi^2}{12}-C^2-3C-7\right)\epsilon_1\epsilon_2\nonumber\\
      &&+\left(\frac{\pi^2}{8}-1\right)\epsilon^2_2
      +\left(\frac{\pi^2}{24}-\frac{C^2}{2}\right)\epsilon_2\epsilon_3~,\label{BS0}\\
b_{s1}&=&n_s-1
      =-2\epsilon_1-\epsilon_2-2\epsilon^2_1-(2C+3)\epsilon_1\epsilon_2-C\epsilon_2\epsilon_3~,\label{BS1}\\
b_{s2}&=&\alpha_s=-2\epsilon_1\epsilon_2-\epsilon_2\epsilon_3~,\label{BS2}
\end{eqnarray}
and those for the tensor spectrum are:
\begin{eqnarray}
b_{t0}&=&-2(C+1)\epsilon_1+\left(\frac{\pi^2}{2}-2C-7\right)\epsilon^2_1
      +\left(\frac{\pi^2}{12}-C^2-2C-2\right)\epsilon_1\epsilon_2~,\label{BT0}\\
b_{t1}&=&n_t=-2\epsilon_1-2\epsilon^2_1-2(C+1)\epsilon_1\epsilon_2~,\label{BT1}\\
b_{t2}&=&\alpha_t=-2\epsilon_1\epsilon_2~,\label{BT2}
\end{eqnarray}
where $C\equiv\ln2+\gamma_E-2\approx-0.7296$ ($\gamma_E$ is the
Euler-Mascheroni constant). And the ratio of amplitudes of the
scalar to the tensor at the pivot scale is:
\begin{eqnarray}
r&=&16\epsilon_1\left[1+C\epsilon_2+\left(C+5-\frac{\pi^2}{2}\right)\epsilon_1\epsilon_2
+\left(\frac{C^2}{2}-\frac{\pi^2}{8}+1\right)\epsilon^2_2+\left(\frac{C^2}{2}-\frac{\pi^2}{24}\right)\epsilon_2\epsilon_3\right].\label{RRR}
\end{eqnarray}
At the first order approximation Eq.(\ref{RRR}) becomes the
well-known consistency relation of inflation $r=-8n_t$.

For dark energy, we choose the commonly used parametrization of
the dark energy equation of state (EoS) as \cite{Linderpara}:
\begin{equation}
\label{Lin} w_{\rm DE}(a) = w_{0} + w_{1}(1-a)~,
\end{equation}
where $a=1/(1+z)$ is the scale factor and $w_{1}=-dw/da$
characterizes the ``running" of the equation of state (RunW
henceforth). For comparison we also consider the $\Lambda$CDM
model and the dark energy model with a constant equation of state
(WCDM henceforth). When using the MCMC global fitting strategy to
constrain the cosmological parameters, it is crucial to include
dark energy perturbation \cite{wmap3:2006:1,Weller:2003hw,Pert}.
However, it is divergent when the parameterized EoS crosses $w=-1$
\cite{Xia:2007km}. By virtue of Quintom dark energy model
\cite{Feng:2004ad}, whose EoS can smoothly cross $w=-1$, the
perturbation at the crossing points is continuous. Thus we have
proposed a technique to treat dark energy perturbation in the
whole parameter space, including $w>-1$, $w<-1$ and at the
crossing points. For details of this method, we refer the readers
to our previous companion papers \cite{Pert}.

In this study, we have modified the publicly available Markov
Chain Monte Carlo package CAMB\footnote{http://camb.info/.}
\cite{camb} / CosmoMC\footnote{http://cosmologist.info/cosmomc/.}
\cite{cosmomc} to include the dark energy perturbation and the
public available code by Leach and Liddle for the primordial
spectrum\footnote{http://astronomy.sussex.ac.uk/$\sim$sleach/inflation/camb${}_{-}$inflation.html/.}
\cite{InfCode}. We assume purely adiabatic initial conditions and
a flat universe. Our most general parameter space is:
\begin{equation}
\mathcal{P}\equiv (\omega_{b}, \omega_{c}, \Theta_{s}, \tau,
w_{0}, w_{1}, \epsilon_1, \epsilon_2, \epsilon_2\epsilon_3,
\log[10^{10}A_s])~,
\end{equation}
where $\omega_{b}\equiv\Omega_{b}h^{2}$ and
$\omega_{c}\equiv\Omega_{c}h^{2}$ are the physical baryon and cold
dark matter densities relative to the critical density,
$\Theta_{s}$ is the ratio (multiplied by $100$) of the sound
horizon to the angular diameter distance at decoupling, $\tau$ is
the optical depth to reionization, $A_{s}$ is defined as the
amplitude of the primordial spectrum. For the pivot scale of the
primordial spectrum we set $k_{\ast} = 0.05$ Mpc$^{-1}$.

In the computation of CMB we have included the WMAP3
Temperature-Temperature (TT) and Temperature-Polarization (TE)
power spectra with the routine for computing the likelihood
supplied by the WMAP team \cite{wmap3:2006:1,wmap3:2006} as well
as the smaller scale experiments, including Boomerang-2K2
\cite{MacTavish:2005yk}, CBI \cite{Readhead:2004gy}, VSA
\cite{Dickinson:2004yr} and ACBAR \cite{Kuo:2002ua}. For the Large
Scale Structure information, we have used the Sloan Digital Sky
Survey (SDSS) luminous red galaxy (LRG) sample
\cite{Tegmark:2006az} and 2dFGRS \cite{Cole:2005sx}. To be
conservative but more robust, in the fitting to the SDSS LRG
sample we have used the first $15$ bins only, $0.0120<k_{\rm
eff}<0.0998$, which are supposed to be well within the linear
regime. For SNIa we have marginalized over the nuisance parameter
\cite{DiPietro:2002cz}. The supernova data we use are the ESSENCE
(192 sample) data \cite{essence}. Furthermore, we make use of the
Hubble Space Telescope (HST) measurement of the Hubble parameter
$H_{0}\equiv 100h$~km~s$^{-1}$~Mpc$^{-1}$ \cite{Hubble} by
multiplying the likelihood by a Gaussian likelihood function
centered around $h=0.72$ and with a standard deviation
$\sigma=0.08$. We also impose a weak Gaussian prior on the baryon
density $\Omega_{b}h^{2}=0.022\pm0.002$ (1 $\sigma$) from Big Bang
Nucleosynthesis \cite{BBN}. Simultaneously we will also use a
cosmic age tophat prior as 10 Gyr $< t_0 <$ 20 Gyr.

For each regular calculation, we run $8$ independent chains
comprising of $150,000-300,000$ chain elements. The average
acceptance rate is about $30\%$. We test the convergence of the
chains by Gelman and Rubin criteria \cite{R-1} and find $R-1$ is
of order $0.01$ which is more conservative than the recommended
value $R-1<0.1$.


\begin{table*}

{\footnotesize

TABLE I. Mean $1\sigma$ constraints on the cosmological parameters
using the current observations. For columns I and II, the
Horizon-flow parameters are obtained at the first and second order
approximation in the slow-roll expansion respectively. For the
weakly constrained parameters, $\epsilon_1$ and $r$, we quote the
$95\%$ upper limit instead.

\begin{center}
\begin{tabular}{l|ccc|ccc}

\hline\hline

&\multicolumn{3}{c|}{I}&\multicolumn{3}{c}{II}\\

Parameter&$\Lambda$CDM&WCDM&RunW&$\Lambda$CDM&WCDM&RunW\\

\hline

$10^2\epsilon_1$&$<1.39$&$<2.18$&$<2.21$&$<2.72$&$<3.10$&$<3.11$\\

$\epsilon_2$&$0.034\pm0.024$&$0.004\pm0.036$&$-0.006\pm0.039$&$0.131\pm0.055$&$0.106\pm0.063$&$0.094\pm0.069$\\

$\epsilon_2\epsilon_3$&$0$&$0$&$0$&$0.089\pm0.042$&$0.080\pm0.043$&$0.072\pm0.044$\\


$n_s$&$0.955\pm0.019$&$0.977\pm0.024$&$0.986\pm0.028$&$0.905\pm0.027$&$0.919\pm0.033$&$0.926\pm0.038$\\

$\alpha_s$&$0$&$0$&$0$&$-0.092\pm0.045$&$-0.082\pm0.045$&$-0.075\pm0.046$\\

$r$&$<0.223$&$<0.349$&$<0.354$&$<0.391$&$<0.457$&$<0.458$\\

$w_0$&$-1$&$-0.888\pm0.064$&$-1.02\pm0.15$&$-1$&$-0.936\pm0.078$&$-1.02\pm0.17$\\

$w_1$&$0$&$0$&$0.474^{+0.509}_{-0.538}$&$0$&$0$&$0.301^{+0.632}_{-0.612}$\\



\hline

$\Delta\chi^2_{\rm min}$&$0$&$-2.0$&$-4.0$&$0$&$-0.2$&$-1.8$\\

\hline\hline

\end{tabular}
\end{center}}
\end{table*}

\section{Results}

\begin{figure*}[htbp]
\begin{center}
\includegraphics[scale=0.65]{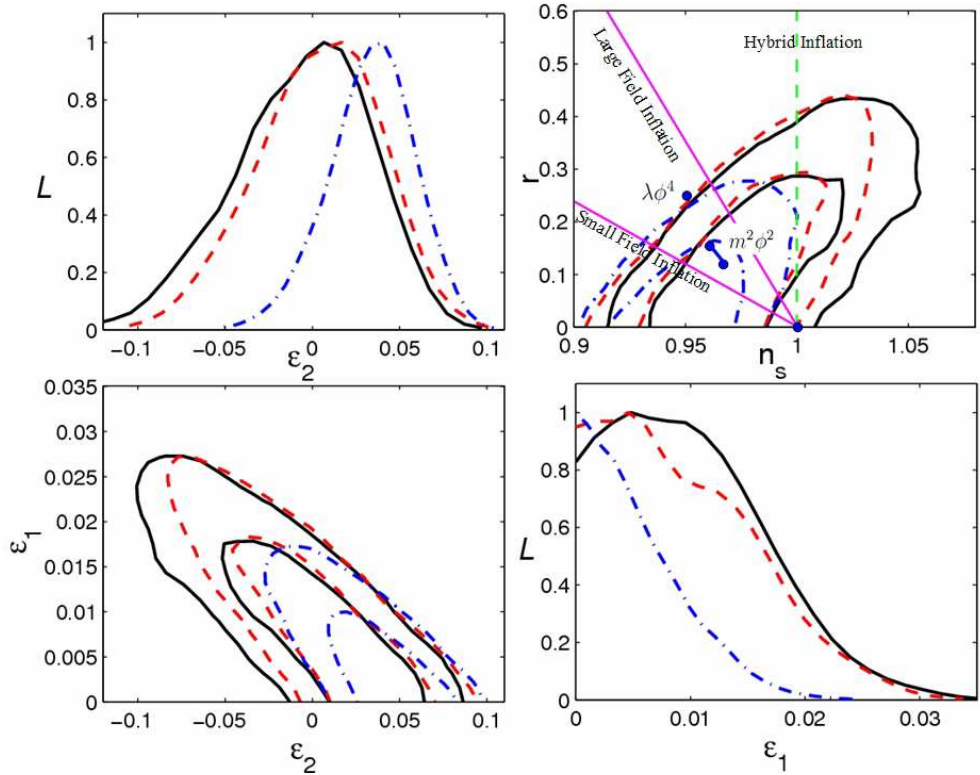}
\caption{Constraints on the inflationary parameters at first order
in the slow-roll approximation in the $\Lambda$CDM (blue
dash-dotted lines), WCDM (red dash lines) and RunW (black solid
lines) dark energy models respectively. The left up panel and
right down panel are the one dimensional marginalized distribution
of Horizon-flow parameters $\epsilon_1$ and $\epsilon_2$. The left
down panel gives the two dimensional constraint on
($\epsilon_2$,$\epsilon_1$). And the right up panel gives the two
dimensional constraint on ($n_s$,$r$). The contours stand for the
$68\%$ and $95\%$ confidence level. The two solid magenta lines
delimit the three classes of inflation models, namely,
small-field, large-field and hybrid models. The blue points are
predicted by $m^2\phi^2$ model and $\lambda\phi^4$ model
respectively with the number of e-foldings, $N$, being $50-60$ for
$m^2\phi^2$ model and $64$ for $\lambda\phi^4$ model.\label{fig1}}
\end{center}
\end{figure*}

We summarize our main global fitting results in Table I. Table I
lists all of the relevant one-dimensional median values and
$1\sigma$ constraints. Shown together are the corresponding
reduction of $\chi^2_{\rm min}$ values compared with the
$\Lambda$CDM model. For the constraints on $\epsilon_1$ and $r$
only $2\sigma$ upper bounds have been shown.

Firstly we consider the first order approximation in the slow-roll
expansion, where the relevant parameters are $\epsilon_1$ and
$\epsilon_2$. In this case the running of the spectral index
vanishes. In the $\Lambda$CDM model, illustrated in the left up
panel and right down panel of Fig.\ref{fig1}, we obtain the
constraints on the Horizon-flow parameters are
$\epsilon_1<0.014~(95\%~C.L.)$ and
$\epsilon_2=0.034\pm0.024~(1\sigma)$ and consequently with
Eq.(\ref{BS1}) and Eq.(\ref{RRR}) we obtain the spectral index
$n_s=0.955\pm0.019~(1\sigma)$ and the tensor-to-scalar ratio
$r<0.223~(95\%~C.L.)$, which is in good agreement with the WMAP's
results \cite{wmap3:2006:1}.

From the right up panel of Fig.\ref{fig1}, one can see that a pure
HZ spectrum for scalar perturbations with no tensors ($n_s=1$,
$r=0$) is clearly disfavored at more than $2\sigma$ by the current
observations. We plot the same constraints in terms of the
Horizon-flow parameters $\epsilon_1$ and $\epsilon_2$ in the left
down panel of Fig.\ref{fig1}\footnote{Ref.\cite{Schwarz:2004tz}
constrained on the dynamics of Inflation in the
($\epsilon_2$,$\epsilon_1$) plane straightforwardly.}. For the
simple monomial chaotic models, the single slow-rolling scalar
field with potential $V(\phi)\sim m^{2}\phi^{2}$, which predicts
$(n_s,r)=(1-2/N,8/N)$, is well within $1\sigma$ region, while
another single slow-rolling scalar field with potential
$V(\phi)\sim \lambda\phi^{4}$, which predicts
$(n_s,r)=(1-3/N,12/N)$, is excluded by more than $2\sigma$ in the
$\Lambda$CDM model \cite{Tegmark:2006az,seljak,eps31,eps32}.

However, the current observational data don't exclude the
dynamical dark energy models and especially mildly favor a class
of models with EoS across the cosmological constant boundary
\cite{Feng:2004ad}. Due to the degeneracy between inflation and
dark energy, it's necessary to perform an analysis of global
fitting allowing simultaneously the dynamics in both inflation and
the dark energy sector.

In the framework of dynamical dark energy models, we find that the
constraints on the Horizon-flow parameters $\epsilon_1$ and
$\epsilon_2$ and the derived parameters $n_s$ and $r$ have been
weakened dramatically as shown in the one dimensional distribution
plots of Fig.\ref{fig1}. Quantitatively, the constraints on the
Horizon-flow parameters are $\epsilon_1<0.022~(95\%~C.L.)$ for
both models and $\epsilon_2=0.004\pm0.036~(1\sigma)$ for WCDM
model, $\epsilon_2=-0.006\pm0.039~(1\sigma)$ for RunW model. And
constraints on the derived parameters are
$n_s=0.977\pm0.024~(1\sigma)$ for WCDM model,
$n_s=0.986\pm0.028~(1\sigma)$ for RunW model, and
$r<0.35~(95\%~C.L.)$ for both models. The mean value of $n_s$ gets
closer to $n_s=1$ and the $95\%$ upper limit of $r$ is relaxed due
to the degeneracy that the tensor fluctuation and the dark energy
component, through the ISW effect, mostly affect the large scale
(low multipoles) TT power spectrum of CMB
\cite{Xia:Inf1,Xia:Inf2,Seljak:2004xh}.

Because of this degeneracy, in the two dimensional plot of
Fig.\ref{fig1}, we can find that the allowed parameter space is
enlarged dramatically. Consequently the HZ spectrum, disfavored
about $3\sigma$ in the $\Lambda$CDM model, can be allowed within
the $2\sigma$ region in the presence of the dynamics of dark
energy. And interestingly many hybrid inflation models, excluded
in the $\Lambda$CDM model, revive in the framework of dynamical
dark energy models as illustrated in Fig.\ref{fig1}
\cite{Xia:Inf2}.

\begin{figure}[htbp]
\begin{center}
\includegraphics[scale=0.6]{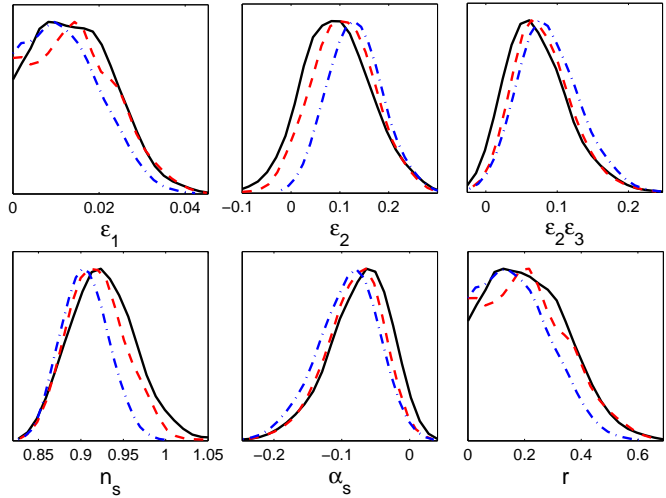}
\caption{Marginalized posterior probability distributions for the
Horizon-flow parameters, $\epsilon_1$, $\epsilon_2$ and
$\epsilon_2\epsilon_3$, and the derived inflationary parameters,
$n_s$, $\alpha_s$ and $r$, up to second order approximation in the
slow-roll expansion in the $\Lambda$CDM (blue dash-dotted lines),
WCDM (red dash lines) and RunW (black solid lines) dark energy
models respectively.\label{fig2}}
\end{center}
\end{figure}

With the second order approximation in the slow-roll expansion,
one has to consider the third Horizon-flow parameter $\epsilon_3$.
In our calculation we choose $\epsilon_2\epsilon_3$ as the basic
parameter directly instead of $\epsilon_3$. Practically this will
make our numerical calculation much more efficient. And we notice
that the second order formalism are valid in the limit of
$\epsilon_2\epsilon_3\ll1$, but not $\epsilon_3\ll1$
\cite{eps31,Makarov:2005uh}. We have also checked with
$\epsilon_3$ as a parameter in the fitting and found the
constraint is very poor \cite{InfCode,eps31,eps32}.

In Fig.\ref{fig2} we show the one dimensional marginalized
posterior probability distributions for the Horizon-flow
parameters, $\epsilon_1$, $\epsilon_2$ and $\epsilon_2\epsilon_3$,
and the derived inflationary parameters, $n_s$, $\alpha_s$ and
$r$, up to second order approximation in the slow-roll expansion.
The constraints on the Horizon-flow parameters become weaken
obviously in the dynamical dark energy models relative to the
$\Lambda$CDM model.

Another result is the appearance of large running of scalar
spectral index which has been found in the literature
\cite{wmap3:2006:1,Easther:2006tv}. From the current observations,
illustrated in Fig.\ref{fig2}, we obtain the fully marginalized
value of $\epsilon_2\epsilon_3$ and $\alpha_s$:
$\epsilon_2\epsilon_3=0.089\pm0.042$, $\alpha_s=-0.092\pm0.045$,
which deviate from zero with more than $2\sigma$, in the
$\Lambda$CDM model. This large value of running $\alpha_s$
violates the inequality:
\begin{equation}
|n_s-1|\gg\left|\frac{\alpha_s}{2}\ln
\left(\frac{k}{k_{\ast}}\right)\right|~,
\end{equation}
for $k$ far away from the pivot scale $k_{\ast}$. One possible
explanation is that current observations are not accurate enough
to determine the scalar spectral index yet. This large running
needs much more observation data, such as PLANCK measurement
\cite{Xia:Inf2}, to verify. Indeed, these constraints may be
affected by the Lyman-$\alpha$ forest data \cite{lya}. However, if
this large value of running $\alpha_s$ can be confirmed by the
future measurement, this would be a great challenge to the
single-field inflation model described by the slow roll expansion
which can not produce this large, negative running of scalar
spectral index $\alpha_s$ \cite{Kosowsky:1995aa}. At that time,
two or more Inflationary history or breaking down the slow roll
expansion would be needed \cite{doubleinflation}.

In the framework of dynamical dark energy models, the large
running of scalar spectral index is still allowed. The constraints
on $\alpha_s$ are $\alpha_s=-0.082\pm0.045$ and
$\alpha_s=-0.075\pm0.046$ for the WCDM and RunW models
respectively. The mean value of $\alpha_s$ slightly shift and the
error bars are unchanged. It seems that the correlation between
dark energy parameters and the running $\alpha_s$ is weak
\cite{Seljak:2004xh,Xia:Inf2}. This weak correlation might be
understood from the distinct effects on CMB TT power spectrum. For
the dark energy parameters $w_0$ and $w_1$, the effect on CMB TT
power spectrum can be somewhat identified with a constant
effective equation of state \cite{weff}:
\begin{equation}
w_{\rm eff}\equiv \frac{\int da\Omega(a)w(a)}{\int da\Omega(a)}~.
\end{equation}
Keeping other cosmological parameters unchanged, the TT power
spectrum will be shifted to larger scalar if the effective EoS
$w_{\rm eff}$ becomes larger. However, the negative running
$\alpha_s$ will suppress the amplitude of the spectrum, which can
not be mimicked by adjusting the dark energy parameters only. This
may be the reason of the weak correlation between the dark energy
parameters and the running of scalar spectral index $\alpha_s$.


\section{Summary}

In this paper we perform an analysis of global fitting on the
inflationary parameters in terms of the Horizon-flow parameters
$\{\epsilon_i\}$ in the presence of dynamical dark energy models
from the current observations. Our analysis shows that the
dynamics of dark energy generally weakens the constraints on
inflationary parameters, due to the degeneracy between dark energy
and inflation parameters.

With the first order approximation in the slow-roll expansion, the
constraints on $\epsilon_1$ and $\epsilon_2$ can be significantly
relaxed and the allowed parameter space of
($\epsilon_2$,$\epsilon_1$), ($n_s$,$r$) panels are enlarged
relative to the $\Lambda$CDM model. Consequently the HZ spectrum
($n_s=1$, $r=0$), disfavored about $3\sigma$ in the $\Lambda$CDM
model, can be allowed within $2\sigma$ in the presence of the
dynamics of dark energy. Interestingly many hybrid inflation
models, especially for models with a ``blue" tilt ($n_{s}>1$),
excluded in the $\Lambda$CDM model, revive in the framework of
dynamical dark energy models.

With the second order approximation in the slow-roll expansion, we
use the parameter $d\epsilon_2/dN=\epsilon_2\epsilon_3$ to do the
calculations instead of $\epsilon_3$. We find that the constraints
on the Horizon-flow parameters become weakened further and the
large running of scalar spectral index is still allowed, in the
framework of dynamical dark energy models. The degeneracy between
the running of the scalar spectral index and the dynamics of dark
energy is weak.


\acknowledgments

We acknowledge the use of the Legacy Archive for Microwave
Background Data Analysis (LAMBDA). Support for LAMBDA is provided
by the NASA Office of Space Science. We have performed our
numerical analysis in the Shanghai Supercomputer Center (SSC). We
thank Yi-Fu Cai, Hong Li, Hiranya Peiris, Yun-Song Piao, Tao-Tao
Qiu, Hua-Hui Xiong, Gong-Bo Zhao for helpful discussions. This
work is supported in part by National Natural Science Foundation
of China under Grant Nos. 90303004, 10533010 and 10675136 and by
the Chinese Academy of Science under Grant No. KJCX3-SYW-N2.


\end{document}